\documentclass[a4paper,twoside]{article}
\newcommand{\affil}[1]{$^{\rm #1}$}
\usepackage[authoryear]{natbib}
\bibpunct{(}{)}{;}{a}{}{,}
\usepackage{graphicx,rotating,amsmath}
\date{}

\newcommand{\kms}{\mbox{km\,s$^{-1}$}}

\def\kms{km\,${\rm s}^{-1}$}
\def\kmn{km\,${\rm s}^{-1}$}

\newcommand{\kmss}{km\ s$^{-1}$ }
\newcommand{\vsini}{$\!${\em v\,}sin{\em i}}
\newcommand{\vsinis}{$\!${\em v\,}sin{\em i} }
\title{\large\bf\flushleft High Resolution Spectroscopy and Spectropolarimetry of some late F-/early G-type sun-like stars as targets for Zeeman Doppler imaging.}
\author{\parbox{\textwidth}{\flushleft
\vspace{-0.5cm}
{\it I.A. Waite\affil{A,D}, S.C. Marsden\affil{B,A}, B.D. Carter\affil{A} , E. Al\'{e}cian\affil{C}, C. Brown\affil{A}, D. Burton\affil{A}, and R. Hart\affil{A}}\\
\vspace{0.4cm}
{\small \affil{A}\,Faculty of Sciences, University of Southern Queensland, Toowoomba, 4350, Australia}\\
{\small \affil{B}\,Centre for Astronomy, School of Engineering and Physical Sciences, James Cook University, Townsville, 4811, Australia}\\
{\small \affil{C}\,LESIA, Observatoire de Paris-Meudon, F-92195 Meudon Cedex, France}                    \\
{\small \affil{D}\,Email: contact waite@usq.edu.au}}} 
\begin{document}
\twocolumn[
\begin{changemargin}{.8cm}{.5cm}
\begin{minipage}{.9\textwidth}
\vspace{-1cm}
\maketitle
%
%
\small{\bf Abstract: High resolution spectroscopy and spectropolarimetry have been undertaken at the Anglo-Australian Telescope in order to identify suitable targets for magnetic studies of young sun-like stars, for the proxy study of early solar evolution. This study involved the investigation of some variable late F-/early G-type sun-like stars originally identified by the Hipparcos mission. Of the 38 stars observed for this study, HIP~31021, HIP~64732, HIP~73780 were found to be spectroscopic binary stars while HIP~19072, HIP~67651 and HIP~75636 are also likely to be binaries {\bf while HIP~33111 could even be a triple system}. Magnetic fields were detected on a number of the survey stars: HIP~21632, HIP~43720, HIP~48770, HIP~62517, HIP~71933, HIP~77144, HIP~89829, HIP~90899 and HIP~105388, making these stars good candidates for follow-up Zeeman Doppler imaging studies.}



\medskip
\medskip
\end{minipage}
\end{changemargin}
]
\small

\section{Introduction}

The study of young sun-like stars provides a window onto the Sun's intensely active past and an understanding of early solar evolution. In particular, the observations of starspots and associated magnetic activity gives clues to the underlying dynamo processes operating in young sun-like stars. Thus, the study of magnetic fields helps us understand the solar interior as well as its atmosphere. A key question in this respect is: how does the young Sun differ in its internal structure and energy transport systems when compared with the modern-day Sun? The technique of Zeeman Doppler imaging (ZDI) \citep{Semel89,Semel93,Donati03} can be used to map the magnetic topologies of rapidly rotating young sun-like stars to help address this question. 

In the Sun today, strong shear forces are formed at the interface between the solid-body rotation of the radiative zone and the differentially rotating convective layer in a region called the tachocline. In the tachocline differential rotation wraps north-south magnetic field lines around the Sun in the direction of rotation and convective motions act to raise the magnetic fields through the convection zone to emerge at the surface. This interaction converts the global poloidal magnetic field to a toroidal field. This effect is known as the ``$\Omega$-effect". The $\alpha$-effect is the reverse process, converting the global toroidal field to the poloidal field. In contrast, studies by \citet{Donati03} of K-dwarf stars and by \citet{Marsden11a,Marsden11b} and \citet{Waite11} of pre-main-sequence G-type stars show regions of azimuthal fields. \citet{Donati03} interpret this in terms of an $\alpha^{2}\Omega$ dynamo process \citep{Brandenburg89,Moss95} being distributed throughout the entire convection zone and close to the surface of the star itself. 

The Sun today undergoes activity cycles in the form of magnetic reversals, but at what stage do these cycles begin during the early evolution of the star? Recent theoretical work by \citet{Brown10} has suggested that young stars undergo ``attempted" field reversals, where the magnetic field begins to break-up, a signature of an impending reversal, only to reinforce again in the original direction. Resolving the origin of the solar dynamo will help us address the more general question of how stellar magnetic cycles develop in young stars, and affect any attendant emerging planetary systems. The search for ZDI targets is thus motivated by the need to study a sample of young Suns to test recent dynamo theory for these stars. 

The initial search for potential ZDI targets by \citet{Waite05} found two pre-main-sequence stars: HD~106506 and HD~141943. ZDI has been used to map the magnetic topologies of HD~106506 \citep{Waite11} and evolution of the magnetic topologies and variations in the surface differential rotation of HD~141943 \citep{Marsden11a,Marsden11b}. This paper is a follow-on from this initial search for late F-/ early G-type stars. Our search specifically aims to measure the projected rotational velocity, \vsini, radial velocity, the level of magnetic and chromospheric activity, and confirm the expected youthful evolutionary status of these stars for studying the origins of the magnetic dynamo in young sun-like stars. 


\section{Observations at the Anglo-Australian Telescope}

\subsection{Selection criteria}

The Hipparcos space mission \citep{Perryman97} has provided a wealth of stellar astrometry, and revealed many previously unknown variable stars. A large number of these unresolved variables \citep{Koen02} are likely to be eclipsing binaries, but some are expected to be active stars with starspot modulation. From the Hipparcos database, late F-/early G-type unresolved variable stars were extracted from the original list of \citet{Koen02}. To reduce the sample to a manageable number of stars, only those sun-like stars with a variability between $\sim$ 0.04 and $\sim$ 0.1 magnitude were selected. If the variability was less than 0.04, the spot activity (if the variability was due to starspots) on the star would unlikely be sufficient to deform the stellar profiles sufficiently for any spatial information to be recovered using the technique of Doppler imaging (DI). Any variation above $\sim$ 0.1 would most likely be a result of a companion star. A final list of 38 stars was compiled for follow-up high-resolution spectroscopy and spectropolarimetry at the Anglo-Australian Telescope (AAT). 

\subsection{Spectroscopy}

High-resolution spectra of 38 late F-/early G- type stars were observed over two nights of Service observations on the 14th of April and 7th of September, 2008 using the using the University College of London \'{E}chelle Spectrograph (UCLES) at the AAT. The EEV2 chip was used with the central wavelength set to 526.8 nm. The 31 lines per mm grating was used with a slit width of 0.73 mm and slit length of 3.18 mm for observations on the 14th April while the slit width was set to 0.74mm and slit length of 3.17mm was set for observations on the 7th September, 2008. This gave an approximate resolution of 50500, extending from order \#84 to order \#129. A journal of the observations is shown in Table \ref{journal_ucles}.

\begin{table*}
\begin{center}
\caption{Journal of observations using UCLES at the AAT.}
\label{journal_ucles}
\begin{tabular}{llccccccc}
\hline 
  HIP 	&  Spectral & UTSTART   &   Exp Time &   S/N$^{a}$   &   v$_{rad}$$^{b}$   &   \vsini$^{b}$   &   EEW$^{c}$ H$\alpha$  &   EqW$^{d}$ Li   \\ 
	&  Type     & 		&   (sec)    & 	     &   \kms	  &   \kms      &   (m\AA)  &   (m\AA) \\
\hline
UTDATE  &   2008, APR 14  &	&      &	&         &	        &  		\\
 23316  &   G5V$^{1}$ & 10:13:49  &  600  	&  77  &  22.8  & $\sim$ 6 &  330$\pm$6  &  198$\pm$7  	\\ 
 27518  &   G3$^{2}$  & 10:26:08  &  400  	&  72  &   5.2  &   $<$5  &  -46$\pm$19 &   $<$5  	\\ 
 31021${^e}$  &   G3V$^{3}$ & 10:48:24  &  400  	&  73  &   -- 	&  --	  &  --   	&   --  	\\ 
 33111${^f}$  &   G5V$^{1}$ & 10:57:02  &  400  	&  79  &   -- 	&  --	  &  --   	&   --  	\\ 
 33699  &   F8V$^{4}$	& 11:05:36  &  400  	&  59  &  29.8  &   $<$5  &  56$\pm$10  &   41$\pm$3  	\\ 
 41688  &   G6IV/V$^{6}$& 11:15:18  &  60  	&  46  & -20.4  &   $<$5  &  62$\pm$10  &   66$\pm$10  	\\ 
 43720  &   G1V$^{1}$	& 11:18:17  &  400  	&  62  &   2.2  &  38     &  400$\pm$28 &   $<$5  	\\ 
 46949  &   G2/3V$^{5}$	& 11:26:48  &  600  	&  64  &  26.8  &   $<$5  &  23$\pm$11  &   68$\pm$3  	\\ 
 48146  &   G6IV/V$^{6}$& 11:39:30  &  600  	&  49  &   0.5  &    $<$5 &  43$\pm$9   &   68$\pm$1  	\\ 
 48770  &   G7V$^{1}$	& 11:52:21  &  1200  	&  67  &  19.6  &     35  &  990$\pm$90 &  234$\pm$5  	\\ 
 60894  &   G0/1V$^{7}$	& 12:14:57  &  600  	&  29  &  29.4  &   $<$5  &  -36$\pm$30 &    $<$5  	\\ 
 62517  &   G0$^{8}$	& 12:27:56  &  400  	&  68  & -25.5  &  52     &  265$\pm$25 &    47$\pm$24  \\ 
 63734  &   F7/8V$^{9}$	& 12:36:18  &  200  	&  97  &   1.3  &  $\sim$  6    &  121$\pm$11 &  141$\pm$6  	\\ 
 63936  &   F8$^{8}$	& 12:42:02  &  600  	&  70  & -10.0  &   $<$5  &  80$\pm$8   &   $<$5  	\\ 
 64732${^e}$  &   F5V$^{6}$	& 12:54:22  &  150  	&  84  &  --    &   --    &  --   	&   --  	\\ 
 66387  &   G0$^{8}$	& 12:59:05  &  600  	&  55  & -24.1  &   $<$5  &  48$\pm$6   &    $<$5  	\\ 
 67651${^g}$  &   F8$^{10}$	& 13:10:25  &  400  	&  57  &  --    &   --	  &  --   	&   --  	\\ 
 68328  &   G0$^{11}$	& 13:20:13  &  600  	&  76  &   8.8  &  $\sim$6	&  850$\pm$45 &  263$\pm$4  	\\ 
 69338  &   G1V$^{6}$	& 13:32:39  &  300  	&  81  &  -8.6  &   $<$5  &   9$\pm$2   &   61$\pm$3  	\\ 
 70053  &   G0$^{8}$	& 13:39:00  &  450  	&  68  &   7.8  &   $<$5  &   17$\pm$7  &   45$\pm$4  	\\ 
 71933  &   F8V$^{1}$	& 13:57:17  &  150  	&  84  &   6.0  &      75 &  274$\pm$25 &  139$\pm$7  	\\ 
        &   		& 15:20:59  &  150  	&  82  &  -1.9  &      75 &    $''$       &   $''$    	\\ 
 71966  &   F7V$^{3}$ 	& 14:01:16  &  450  	&  75  &   9.3  &   $<$5  &   90$\pm$10 &   30$\pm$3  	\\ 
 73780${^e}$  &   G0IV/V$^{7}$& 14:10:35  &  300  	&  77  &  --    &   --    &  --   	&   --  	\\ 
 75636${^g}$  & G9V$^{1}$&   14:17:19  &  900  	&  88   &  43.7 &  50   &  820$\pm$90  &   $<$5  \\ 
 77144  &   G1V$^{11}$	& 14:33:28  &  450  	&  93  &   -1.6 &  65     &  466$\pm$24 &  207$\pm$7  	\\ 
 79090  &   F8$^{8}$	& 14:43:39  &  600  	&  60  &   6.1  &   $<$5  &  56$\pm$12  &   46$\pm$4  	\\ 
 79688  &   G1V$^{5}$	& 14:56:07  &  200  	&  86  &  12.6  &  11     &  172$\pm$12 &   13$\pm$5  	\\ 
 89829  &  G5V$^{1}$ 	& 15:01:17  &  200  	&  74  & -10.6  &  114    &  280$\pm$52 &  211$\pm$13 	\\ 
        &   		& 15:25:42  &  200  	&  75  &   1.2  &  114    &   $''$        &   $''$  	\\  
 90899  &  G1V$^{12}$ 	& 15:06:00  &  450  	&  82  &  -2.7  &  19     &  408$\pm$14 &  176$\pm$6  	\\ 
 93378  &  G5V$^{1}$ 	& 15:16:07  &  150  	&  76  &     -- &  225    &  0  	&  322$\pm$56 	\\ 
  	&   	& 15:30:48  &  150  	&  68  &     -- &  229    &  $''$  	&    $''$ 	\\ 
  	&   	& 16:11:01  &  200  	&  62  &     -- &  226    &  $''$  	&    $''$ 	\\ 
105388  &   G7V$^{1}$	& 16:04:01  &  300  	&  58  &  -1.8  &  17     &  520$\pm$50 &   216$\pm$5 	\\ 
\hline 
UTDATE  &   2008, SEPT 7  &     &      &	 &	    &       &  	 		\\
  5617  &   G2/3$^{7}$	& 13:49:05  &  400  	&  70  &  56.5  &   $<$5    &  --   &   $<$5  		\\ 
 10699  &   G7IV$^{1}$	& 13:58:52  &  600  	&  82  &  39.1  &  7        &  220$\pm$20  &   32$\pm$3 \\ 
 11241  &   F8V$^{9}$	& 14:13:43  &  400  	&  83  &  -3.5  &   $<$5   &  164$\pm$8    &   98$\pm$3 \\ 
 19072$^{f}$  & F8$^{8}$& 18:19:56  &  240  	&  79  &  --    &   --     &  --   	   &   --  		\\ 
 	&   		& 18:56:21  &  240  	&  71  &  --    &   --     &  --   &   --  		\\ 
 20994  &   G0$^{8}$	& 18:25:13  &  900  	&  74  &  57.8  &   $<$5   &   85$\pm$31   & 44$\pm$2   \\ 
 21632${^h}$  &   G3V$^{1}$ & 14:23:12  &  400  &  81  &  19.5  &  18      &  385$\pm$21  &  190$\pm$2  \\ 
 	&   		& 19:04:37  &  400  	&  97   & 19.3  &  18      &  517$\pm$32  &  $''$ 	\\ 
 25848  &  G0$^{13}$ 	& 18:42:35  &  600  	&  68  &  27.3  &  69      &  668$\pm$60  &  250$\pm$13 \\ 
 	&   		& 19:13:43  &  600  	&  65  &  28.7  &  69      &   $''$         &  $''$  	\\ 

\hline
\hline

\end{tabular}
\end{center}
$^{a}$S/N: Mean Signal-to-noise at Order 107, which was the centre of the spectrum. $^{b}$The radial velocity (v$_{rad}$) and projected rotational velocity (\vsini). The errors are estimated to be $\pm$1 \kms, although for rapidly rotating stars with substantial deformation of the line profiles due to spot features, the errors could increase to $\pm$3 \kmss or higher. $^{c}$EEW H$\alpha$: Emission equivalent width of the H$\alpha$ line, see Section 3.3. $^{d}$EqW Li: Equivalent width for the Li-670.78 nm spectral line. ${^e}$Binary system. ${^f}$Possible triple system. $^{g}$Possible binary system. $^{h}$ This star has shown variation in the H$\alpha$ profile hence both measurements for the emission equivalent width have been given. \\
$^{1}$\citet{Torres06}, $^{2}$\citet{Jackson54}, $^{3}$\citet{Houk75}, $^{4}$\citet{Rousseau96}, $^{5}$\citet{Houk82}, $^{6}$\citet{Houk99}, $^{7}$\citet{Houk78}, $^{8}$\citet{SAO66}, $^{9}$\citet{Houk88}, $^{10}$\citet{Dieckvoss75}, $^{11}$\citet{Sartori03}, $^{12}$\citet{Turon93}, $^{13}$\citet{Li98}\\
\medskip\\
\end{table*}

\subsection{Spectropolarimetry}
\label{sempol_obs}
Follow-up spectropolarimetric observations of stars that exhibited rapid rotation were undertaken on a number of Director's nights at the AAT using the Semel Polarimeter (SEMPOL) \citep{Semel89,Semel93,Donati03}. Again the EEV2 chip was used, with a central wavelength of 526.8nm and coverage from 437.6497 nm to 681.8754 nm. The dispersion of $\sim$ 0.002481 nm at order \# 129, with an average resolution across the chip of approximately $\sim$ 70 000. Observations in circular polarisation (Stokes {\it V}) consists of a sequence of four exposures. After each of the exposures, the half-wave Fresnel Rhomb of the SEMPOL polarimeter was rotated between +45$^o$ and -45$^o$ so as to remove instrumental polarisation signals from the telescope and the polarimeter. Section \ref{spectro_sect} gives more details regarding spectropolarimetric observations while more details on the operation of SEMPOL is given in \citet{Semel93}, \citet{Donati97} and \citet{Donati03}. A journal of the observations is shown in Table \ref{journal_sempol}.

\begin{sidewaystable*}[ht]
\begin{center}
\caption{Journal of Spectropolarimetric observations using SEMPOL at the AAT}
\label{journal_sempol}
\begin{tabular}{lcccclc}
\hline	
HIP    & UTDATE & UT Time$^a$	  & Exposure Time  & Mean S/N$^c$     & Magnetic   & FAP$^d$	\\
number & 	& 	          & (seconds) $^b$ & (Stokes $\it{V}$)       & Detection? &            \\
\hline	     
21632  & 2008 Dec10 & 12:39:15 & 4 x 900 & 5898 &  No Detection		& 2.245 x 10$^{-01}$ \\
21632  & 2008 Dec13 & 14:27:53 & 2 x 900 $^e$ & 2710 & Definite 	& 4.761 x 10$^{-06}$ \\
43720  & 2008 Dec09 & 16: 4:50 & 4 x 900 & 5304 & Definite 		& 0.000		     \\
43720  & 2008 Dec14 & 14:12:40 & 4 x 900 & 2626 & No Detection 		& 1.667 x 10$^{-02}$ \\
43720  & 2009 Apr09 & 11: 8:13 & 4 x 800 & 3800 & Definite 		& 5.194 x 10$^{-10}$ \\
43720  & 2009 Apr09 & 12:58:33 & 4 x 800 & 3759 & Definite 		& 6.871 x 10$^{-12}$ \\
48770  & 2009 Dec03 & 16:42:47 & 4 x 750 & 1734 & No Detection		& 1.528 x 10$^{-02}$ \\
48770  & 2010 Apr01 & 10:30:40 & 4 x 800 & 2530 & Definite 		& 7.011 x 10$^{-10}$ \\
62517  & 2009 Apr09 & 15: 7:23 & 4 x 800 & 1559 & No Detection 		& 8.504 x 10$^{-01}$ \\
62517  & 2010 Apr02 & 12: 0:30 & 4 x 800 & 6685 & Marginal Detection 	& 1.786 x 10$^{-03}$ \\
71933  & 2008 Dec18 & 17:44:15 & 4 x 900 & 1748 & No Detection 		& 7.146 x 10$^{-01}$ \\
71933  & 2010 Apr01 & 15:37:08 & 4 x 800 & 8843 & Definite		& 6.457 x 10$^{-11}$ \\
77144  & 2010 Mar31 & 18:02:03 & 4 x 800 & 4315 & Definite 		& 0.000		     \\
77144  & 2010 Apr03 & 17:54:31 & 4 x 800 & 3783 & No Detection 		& 1.362 x 10$^{-01}$ \\
89829  & 2009 Apr13 & 17:10:25 & 4 x 600 & 2544 & No Detection 		& 1.432 x 10$^{-01}$ \\
89829  & 2010 Apr01 & 16:41:14 & 4 x 800 & 7274 & Definite  		& 4.663 x 10$^{-15}$\\
90899  & 2010 Mar28 & 17:50:20 & 2 x 800$^e$ & 669 & No Detection	& 6.837 x 10$^{-01}$ \\
90899  & 2010 Apr02 & 17:42:47 & 4 x 800 & 3663	& Marginal Detection    & 3.647 x 10$^{-3}$ \\
93378  & 2010 Apr01 & 17:43:42 & 4 x 800 & 9727	& No Detection	 	& 7.102 x 10$^{-2}$ \\
105388 & 2008 Dec09 & 10:17:50 & 4 x 900 & 3822 & Definite    	 	& 1.732 x 10$^{-14}$\\ 
105388 & 2008 Dec10 & 10:23:32 & 4 x 900 & 3388 & No Detection 		& 5.084 x 10$^{-01}$\\

\hline
\end{tabular}
\medskip\\
$^a$ Mid-observing time  \\
$^b$ Generally a cycle consists of four sub-exposures.		\\
$^c$ Mean Signal-to-noise in the Stokes $\it{V}$ LSD profile, see Section \ref{Data_Analysis}.  \\ 
$^d$ FAP: False Alarm Probability. See Section \ref{spectro_sect} for more details.	\\
$^e$ Due to cloud, this cycle was reduced to two sub-exposures. \\
\end{center}
\end{sidewaystable*}

\subsection{Data Analysis}
\label{Data_Analysis}
The aim of this project is to determine the projected rotation velocity (\vsini), radial velocity, level of magnetic and chromospheric activity and estimate the age of each of the targets. The chromospheric activity indicators included the H$\alpha$, magnesium triplet and sodium doublet spectral lines. The Li{\sc i} 670.78 nm spectral line was used as an age indicator. The initial data reduction was completed using the ES{\sc p}RIT software package (\'{E}chelle Spectra Reduction: An Interactive Tool, \citep{Donati97}). The technique of Least Squares Deconvolution (LSD) \citep{Donati97} was applied to each spectra. LSD assumes that each spectral line in the spectrum from a star can be approximated by the same line shape. LSD combines several thousand weak absorption lines to create a high signal-to-noise (S/N) single-line profile. Whereas the average S/N of a typical profile was $\sim$ 50-100, the resulting LSD profile has a combined S/N of the order of 1000 or higher. This substantial multiplex gain has the advantage of removing the noise inherent in each line profile while preserving Stokes $\it{I}$ and $\it{V}$ signatures. The line masks that were used to produce the LSD profile were created from the Kurucz atomic database and ATLAS9 atmospheric models \citep{Kurucz93} and were closely matched to the spectral type of each individual star.

\section{Results and Analysis}

\subsection{Projected Rotational Velocity}
The projected rotational velocity, \vsini, was measured by rotationally broadening a solar LSD profile to match the LSD profile of the star. This method was shown to be reliable, particularly with rapidly rotating stars, by \citet{Waite05} when they compared this technique with the Fast Fourier Transform technique of \citet{Gray92}. Table \ref{journal_ucles} shows the projected rotational velocities for the target stars. The error bars on each measurement are usually $\pm$1 \kms, however, for rapidly rotating stars with substantial deformation of the line profiles due to spot features, the errors could increase to $\pm$3 \kmss or higher. Many of these \vsinis values have not been previously determined. 

The term Ultrafast Rotator (UFR) has been used extensively in the literature but without an explicit definition being applied. Hence the need to refine this terminology, particularly for solar-type stars. Stars that have projected rotational velocities less than 5 \kmss will be considered as Slow Rotators (SR) as this is the lower limit at which we can accurately measure the \vsini\ of the star in this dataset. Those stars with \vsinis between 5 \kmss and  20 \kmss will be considered as Moderate Rotators (MR). This upper limit is considered a critical velocity where dynamo saturation has been theorised to slow the angular momentum loss of rapidly rotating stars (e.g. \citet{Iwrin07, Krishnamurthi97, Barnes96}). Below 20 \kms, the strength of the star's magnetic dynamo is related to the star's rotation rate but above this speed, it is believed that dynamo saturation is occurring where the strength of the star's magnetic dynamo is no longer dependent upon stellar rotation. One empirical measure of this saturation in young solar-type stars is coronal X-ray emission. This emission, defined as the ratio of the star's X-ray luminosity to that of the star's bolometric luminosity \citep{Vilhu84}, increases as rotation increases when it plateau's at the \vsinis of 20 \kms. \citet{Stauffer97} theorised that this is consistent with dynamo saturation. Those stars with \vsinis greater than 20 \kmss to 100 \kmss will be referred to as Rapid Rotators (RR). The upper limit of $\sim$ 100\kmss was selected as at this rotational speed, the X-ray luminosity decreases below the saturated level \citep{Randich98}, an effect that \citet{Prosser96} called supersaturation. These definitions are consistent with those used by \citet{Marsden09}, \citet{Marino03} and \citet{Prosser96}. Those stars between 100 \kmss to 200 \kmss will be referred to as Ultra-Rapid Rotators (URR). Stars that exceed 200 \kmss will be referred to as Hyper-Rapid Rotators (HRR) and are likely to be very oblate. However many of the stars in this sample have not had their inclination determined, thus the \vsinis is likely to be an underestimate of the true equatorial rotation of the star, and thus depending on the inclination, the star may be more rapidly rotating than indicated by the \vsini. In addition, we limit these definitions to solar-type stars. Table \ref{vsini_def} gives a summary of the definitions used in this paper.

\begin{table}
\begin{center}
\caption{Classification of Solar-type stars based on projected rotational velocities.}
\label{vsini_def}
\begin{tabular}{lc}
\hline	
Classification & \vsini~ range  \\
               & (\kms)          \\
\hline
Slow Rotator (SR)         & 0 - $<$ 5      \\
Moderate Rotator (MR)     & 5 - $<$ 20     \\ 
Rapid Rotator (RR)        & 20 - $<$ 100   \\
Ultra-Rapid Rotator (URR) & 100 - $<$ 200   \\
Hyper-Rapid Rotator (HRR) & 200+         \\  
\hline
\end{tabular}
\medskip\\
\end{center}
\end{table}

\subsection{Heliocentric Radial Velocity}
\label{vrad}
Each spectrum, when extracted, was shifted to account for two effects. Initially, small instrumental shifts in the spectrograph were corrected for by using the positions of the telluric lines embedded in each spectrum. A telluric line mask was used to produce an LSD profile and the exact position of this profile was used to determine these small corrections \citep{Donati03}. Secondly, the heliocentric velocity of the Earth towards the star was determined and corrected for. 

The LSD profile of the star was used to measure the radial velocity of the star by first re-normalising the profile, then fitting a gaussian curve to the profile and measuring the location of the minimum. The radial velocities are listed in Table \ref{journal_ucles}. The error in the radial velocity was estimated to be $\pm$1 \kmss although the presence of spots on the surface did have some effect on the location of the minimum, especially on stars with rotational speeds in excess of 100 \kmss such as HIP~89829 where the fitting of the gaussian profile was problematic given the rapid rotation. The large variation seen in the radial velocity measurements for HIP~89829 could be a result of this star being a binary star. However, we have searched for evidence of a companion star deforming the LSD profile of the spectropolarimetry data and conclude that this star is probably single. This still does not preclude the existence of a secondary component as a very low mass companion such as an M-dwarf is likely to modify the radial velocity of the primary without generating a detectable line in the LSD profile, because the line-mask employed here is optimised for the primary, not for the companion. It was impossible to accurately measure the radial velocity of the HRR HIP~93378 due to its extreme deformation of the LSD profiles.

\subsection{Chromospheric Indicators: Hydrogen $\alpha$, Magnesium-I triplet and Sodium-I doublet. }

\begin{figure}[ht] 
\begin{center}
\includegraphics[scale=0.50, angle=0]{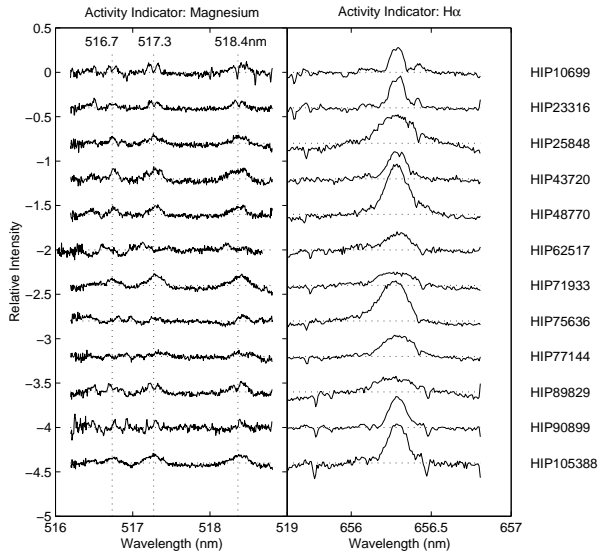}
\caption{The core emission in the magnesium triplet lines, left panel, and the H$\alpha$ spectral line, right panel, are shown for rapid rotators in this sample. This core emission was obtained by subtracting a radial velocity corrected, rotationally-broadened solar spectrum. }
\label{activity_indicators}
\end{center}
\end{figure}

The H$\alpha$, magnesium-I triplet (516.733, 517.270 and 518.362 nm) and sodium D$_{1}$ and D$_{2}$ doublet (588.995 and 589.592 nm) lines are often used as a proxies for stellar activity and in particular chromospheric activity. The H$\alpha$ line is formed in the middle of the chromosphere \citep{Montes04} and is often associated with plages and prominences. The magnesium I triplet and sodium D$_{1}$ and D$_{2}$ doublet lines are collisionally dominated and are formed in the lower chromosphere and upper photosphere. This makes them good indicators of changes in that part of the atmosphere of stars \citep{Montes04}. Many authors (e.g. \citet{Zarro83}, \citet{Young89}, \citet{Soderblom93a}, \citet{Montes04}, \citet{Waite05}) determine the emission component of the H$\alpha$ line by subtracting the stellar spectrum from a radial velocity-corrected, inactive star that has been rotationally broadened to match the \vsinis of the target. This technique is temperature dependent, but since all of our targets have similar spectral types, a solar spectrum was used as the inactive standard star in this survey. The resulting emission equivalent width (EEW) is a measure of the active chromospheric component of the these spectral lines, including the H$\alpha$ line. Figure \ref{activity_indicators} shows the core emission of the magnesium triplet and H$\alpha$ line for the more active stars in this sample. Many of the likely targets for ZDI exhibit core emissions; however, the HRR star HIP 93378 exhibits no activity in either the magnesium triplet or the H$\alpha$ line. This may be due to the extreme broadening of the spectral lines ``washing out" the emission component.

\subsection{Lithium-I 670.78-nm: An Age Indicator}
\label{lithium_paragraph}

In the absence of a companion star, an enhanced Li{\sc i} 670.78 nm line can be used as an indicator of youth \citep{Martin96} for stars that are cooler than mid-G type (0.6 $\le$ B-V $\le$ 1.3) although this is not as useful for F-type stars where there appears to be a plateau in the depletion of the lithium due to age \citep{Guillout09}. However, \citet{doNascimento10} point out that there is a large range in lithium depletion for solar-type stars which may reflect different rotational histories or as a result of different mixing mechanisms such as shear mixing caused by differential rotation \citep{Bouvier08}. In analysing the spectra, the equivalent width EqW$_{Li}$ was measured using the {\sc SPLOT} task in {\sc IRAF}. This was done so as to allow comparison with those measured by \citet{Torres06}. The error in the measured values of EqW$_{Li}$ is primarily due to uncertainties in the continuum location. When the rotational velocity, \vsini, exceeded 8 \kms, the Li{\sc i} spectral line is blended with the nearby 670.744 nm Fe{\sc i} line. This was corrected using the same correction factor developed by \citet{Soderblom93a,Soderblom93b}. The correction used is shown in equation \ref{lithium}.
\begin{equation}
\label{lithium}
	EW_{Li}\it{corr} = EW_{Li} - 20 (B-V) - 3
\end{equation}

\begin{figure}[ht] 
\begin{center}
\includegraphics[scale=0.45, angle=0]{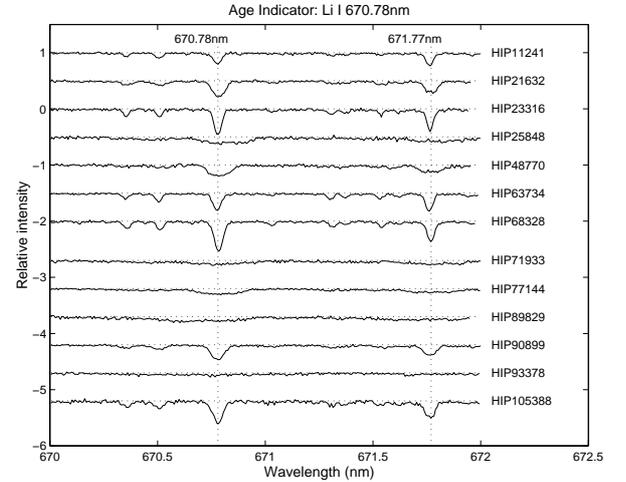}
\caption{The lithium-I 670.78 nm line, compared with the calcium-I 671.77 nm line, for a number of stars in this survey.}
\label{age_indicator}
\end{center}
\end{figure}

Some of the results appear slightly discrepant when compared with \citet{Torres06}. It is unclear whether \citet{Torres06} corrected for the Fe{\sc i} line in their measurements but such a difference in processing may possibly explain the discrepancies. Figure \ref{age_indicator} shows the strength of the Li{\sc i} 670.78 nm line, compared with the nearby Ca{\sc i} 671.80 nm line, for a number of stars that are likely to be future targets for the ZDI programme.

\subsection{Spectropolarimetry}
\label{spectro_sect}

The magnetic signatures embedded in starlight are extremely difficult to detect. The typical Zeeman signature is very small, with a circular polarisation signature (Stokes {\it V}) of $\sim$0.1\% of the continuum level for active stars \citep{Donati97}. As discussed in Section \ref{sempol_obs}, observations in Stokes {\it V} consists of a sequence of four sub-exposures with the half-wave Fresnel Rhomb being rotated between +45$^o$ and -45$^o$. To detect these signatures, LSD is applied to increase the signal-to-noise of the signature when creating the Stokes $\it{V}$ profile. The Stokes $\it{V}$ profile is the result of constructively adding the individual spectra from the four exposures by ``pair-processing" sub-exposures corresponding to the opposite orientations of the half-wave Fresnel Rhomb. To determine the reliability of the process, a ``null" profile is produced as a measure of the noise within the LSD process. This null profile is found by ``pair-processing" sub-exposures corresponding to the identical positions of the half-wave Fresnel Rhomb of the SEMPOL polarimeter during each sequence of 4 sub-exposures. 

When producing figures, such as Figure \ref{hip21632_lsd} for example, the Stokes $\it{V}$ (upper) and null (middle) profiles have been multiplied by 25 so as to show the variation within each profile. Both profiles have been shifted vertically for clarity. The dots on the Stokes $\it{V}$ and null profiles are the actual data while the smooth curve is a 3-point moving average. The deformation in the Stokes $\it{V}$ profile is a direct result of the magnetic field observed on the star while deformation in the Stokes $\it{I}$ (intensity) profile (lower) is a result of starspots on the surface of the star. For more information on ZDI see \citet{Carter96} and \citet{Donati97}. 

For each observation a false-alarm probability (FAP) of magnetic field detection was determined. FAP is a measure of the chance of the signal found in the Stokes $\it{V}$ profile being a result of noise fluctuations rather than a real magnetic detection. The FAP is based on a $\chi^2$ probability function \citep{Donati92} and is estimated by considering the reduced $\chi^{2}$ statistics both inside and outside the spectral lines, as defined by the position of the unpolarised LSD profiles, for both the Stokes $\it{V}$ and the null profiles \citep{Donati97}. The FAP for each observation is listed in Table \ref{journal_sempol}. A definite magnetic detection in the Stokes $\it{V}$ was considered if the associated FAP was smaller than $10^{-5}$ (i.e. $\chi^2$ probability was larger than 99.999 \%) while a marginal detection was observed if the FAP was less than $10^{-3}$ but greater than $10^{-5}$. In addition to this, the signal must only have been detected in the Stokes $\it{V}$ profile and not within the null profile, and be within the line profile velocity interval, from $v_{rad}$-\vsinis to $v_{rad}$+\vsini. This criteria is consistent with the limits used by \citet{Donati97}.


\section{Some comments on individual stars}

Many of the stars in this survey exhibited some indication of activity either due to the presence of a companion star, or in the case of single stars, activity due to youth and/or rapid rotation. We will consider each likely ZDI target in more detail in this section.

\subsection{Moderate and rapid rotators suitable for ZDI studies}
\subsubsection{HIP~21632}  

\begin{figure}[ht] 
\begin{center}
\includegraphics[scale=0.450, angle=0]{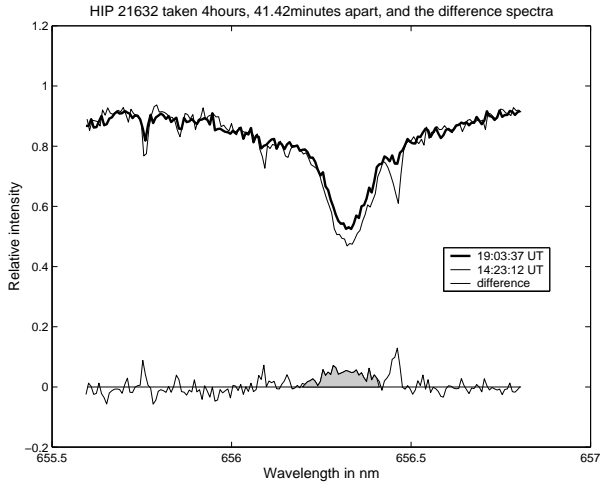}
\caption{The variation of the H$\alpha$ profile of HIP~21632. Two exposures were taken, separated by 4 hours, 41.41 minutes. The variation in the H$\alpha$ profile is shown as a shaded region in the difference spectrum. The sharp absorption line at $\sim$656.4 nm in the 12:23:12UT spectra is most likely due to telluric lines at 656.4049 and 656.4200 nm. }
\label{hip21632_comparisonHA}
\end{center}
\end{figure}

HIP~21632 is a G3V star \citep{Torres06}. The Hipparcos space mission measured a parallax of 18.27$\pm$1.02 milli-arcseconds (mas) \citep{vanLeeuwen07}, giving a distance of $178_{-9}^{+11}$ light-years (ly). Using the bolometric corrections of \citet{Bessell98} and the formulations within that paper, the effective temperature of this star was determined to be 5825$\pm$45 K and the radius was estimated to be 0.96$\pm$0.04 R$_\odot$. The luminosity was subsequently estimated to be $0.93_{-0.11}^{+0.12}$ L$_\odot$. \citet{Zuckerman04} proposed that HIP~21632 was a member of the Tucana/Horologium Association indicating an age of $\sim$30 Myr. This star has an emission equivalent width for the H$\alpha$ line in the range from $\sim$ 385 m\AA\ to $\sim$ 517 m\AA\ demonstrating the presence of a very active, and variable, chromosphere. This variation could be due to prominences occuring on this star. Two exposures, separated by 4 hours 41.41 minutes, were taken on April 14, 2008 demonstrated noticeable variation in the core emission of the H$\alpha$ line, as shown in Figure \ref{hip21632_comparisonHA}. Yet there was no variation in the magnesium triplet and indeed, there was very little core emission in the three lines, with some filling-in in the wings of the lines. However, there were some minor changes in the sodium D$_{1}$ line but the variation is no where near as pronounced as in the mid-level chromospheric level. There is no evidence of the presence of a companion star in the LSD profile. \citet{Torres06} measured a radial velocity of 18.8 \kmss (using cross correlation methods) and an equivalent width for the Li{\sc i} line of 200 m\AA. These values are consistent with the values obtained from this survey of an average radial velocity of 19.4$\pm$1 \kmss and a equivalent width for the Li{\sc i} line of 190$\pm$2 nm (see Section \ref{lithium_paragraph} on a possible reason for slightly discrepant values). 

\begin{figure}[ht] 
\begin{center}
\includegraphics[scale=0.3, angle=0]{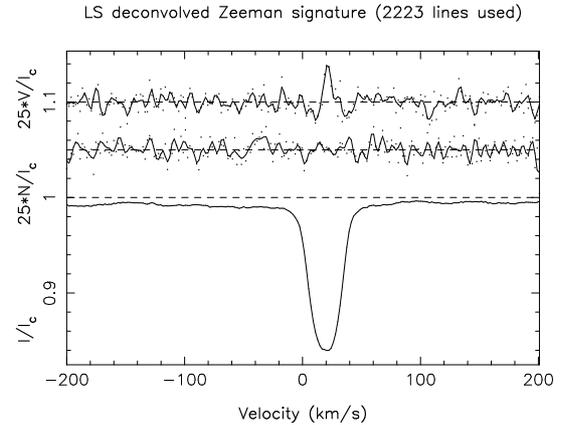}
\caption{The magnetic detection of young G3V star HIP~21632. The lower profile is the Stokes $\it{I}$ LSD profile, the middle profile is the null profile and the upper profile is the Stokes $\it{V}$ profile. The dots are the actual data while the smooth line is a moving 3-point average of the data. The Stokes $\it{V}$ and Null profiles have been vertically shifted for clarity. In addition to this, the Stokes $\it{V}$ and Null profiles each were multiplied by 25 in order to highlight the actual signatures. The Stokes $\it{V}$ profile clearly shows a strong magnetic detection. This was achieved using only 2 sub-exposures with a mean S/N of 2560.}
\label{hip21632_lsd}
\end{center}
\end{figure}

This star was observed spectropolarimetrically on two occasions using SEMPOL. On the first occasion, a no magnetic signal was detected with a mean S/N of 5898 with 4 sub-exposures, yet achieved a definite magnetic detection with only 2 sub-exposures with a mean S/N of 2710 on the second occasion. The resulting LSD profile for this cycle is shown in Figure \ref{hip21632_lsd}. Whereas this star is only a moderate rotator, it is a worthwhile target for more detailed spectropolarimetric studies.

\subsubsection{HIP~25848} 

\begin{figure*}[ht]
\begin{center}
\includegraphics[scale=0.800,angle=0]{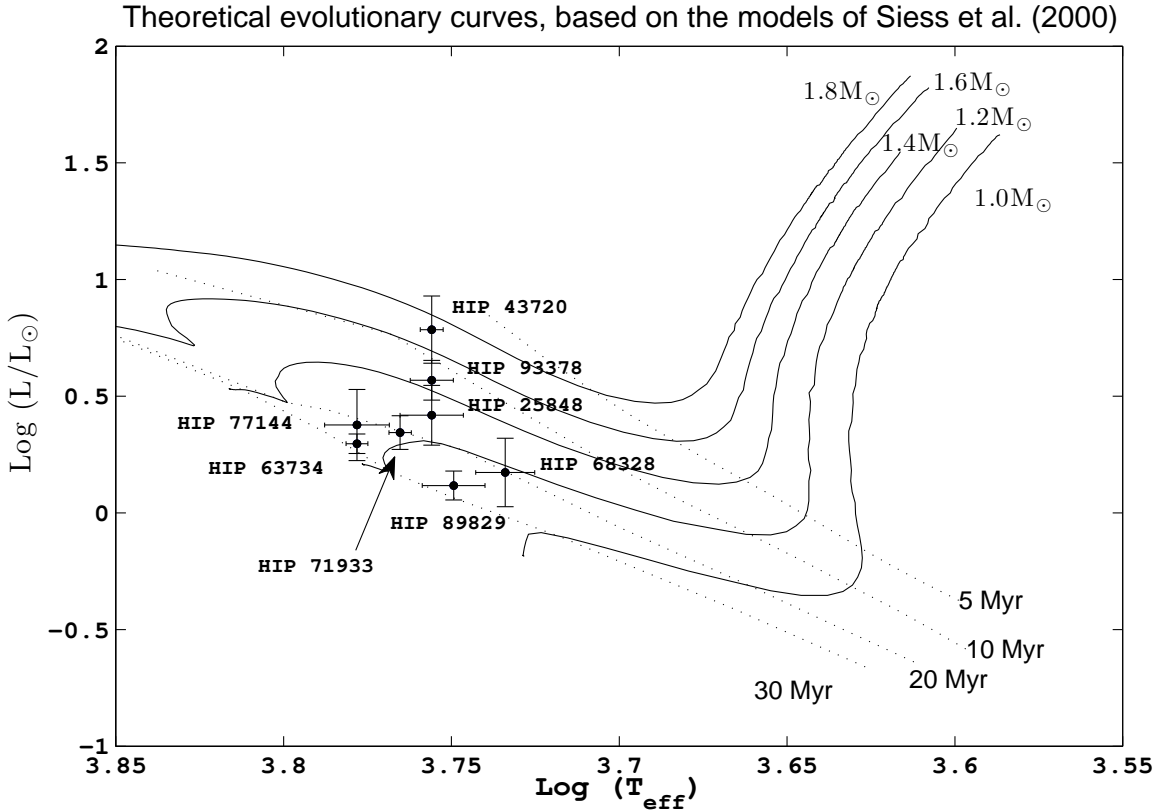}
\caption{The evolutionary status of some of the survey stars, based on the theoretical isochrones of \citet*{Siess00}. Only those likely ZDI targets with accurate photometry were incorporated onto this diagram.}
\label{evolution}
\end{center}
\end{figure*}

HIP~25848 is a G0 weak-lined T Tau-type star \citep{Li98}. It has a trigonometric parallax of 7.95$\pm$1.29 mas \citep{vanLeeuwen07} giving a distance of $410_{-57}^{+79}$ ly. Using the bolometric corrections by \citet*{Bessell98}, the effective temperature was estimated to be 5700$\pm130$ K. Using the formulation contained in \citet*{Bessell98} the star is estimated to be $\sim$ $1.67_{-0.17}^{+0.23}$ R$_\odot$. Placing this star on the theoretical isochrones of \citet{Siess00}, it is estimated to be 1.3$\pm$0.1 M$_\odot$ with an age between 10Myr to 20Myr. This is shown in Figure \ref{evolution}. This is consistent with the age found by \citet{Tetzlaff11}, however slightly less massive than that quoted in that paper. \citet{Norton07} used SuperWASP to measure a period of 0.9426 d. This survey measured a \vsinis of 69 \kms. This star has an emission equivalent width for the H$\alpha$ line of 668$\pm$60 m\AA, meaning that it is very active and one of the most active stars from this survey. The sodium doublet lines were filled in, almost to the continuum. It has a very deep Li{\sc i} line suggesting, in the absence of a companion star, a youthful star. No spectropolarimetric observations were obtained were obtained. With a declination of +23$^{o}$, it would be a difficult target for ZDI at the AAT. However, this star would be an ideal target for ESPaDOnS at the CFHT (Canada-France-Hawaii Telescope, Hawaii) or NARVAL at the TBL (T\'{e}lescope Bernard Lyot, Pic du Midi, France).

\subsubsection{HIP~43720}

HIP~43720 is a particularly active, G1V star \citep{Torres06}. It has a trigonometric parallax of 5.38$\pm$0.94 mas \citep{vanLeeuwen07} giving a distance of $606_{-90}^{+128}$ ly. Using the star's V-I value and the formulation in \citet*{Bessell98}, the star's temperature was estimated to be $5700_{-45}^{+40}$ K while its radius was estimated to be $2.6_{-0.4}^{+0.7}$ R$_\odot$. Placing HIP~43720 onto the theoretical isochrones of \citet*{Siess00}, as shown in Figure \ref{evolution}, suggests that this star's age is $\le$ 10 Myr years and has a mass of between 1.6 and 1.8 M$_\odot$. However, this age estimate is not supported by the depth of the Li{\sc i} line, with an equivalent width of $<$5 m\AA. One can speculate that the lithium has already been depleted. \citet{Guillout09} suggest that stars with deep convective envelopes, such as M-dwarfs, are very efficient at depleting the lithium concentration. Being a pre-main-sequence star, HIP~43720 may also possess a very deep convective zone. Alternatively, \citet{Bouvier08} suggest that this depletion may be due to large velocity shear at the base of the convective zone as a result of star-disk interaction. 

\begin{figure}[ht] 
\begin{center} 
\includegraphics[scale=0.35, angle=0]{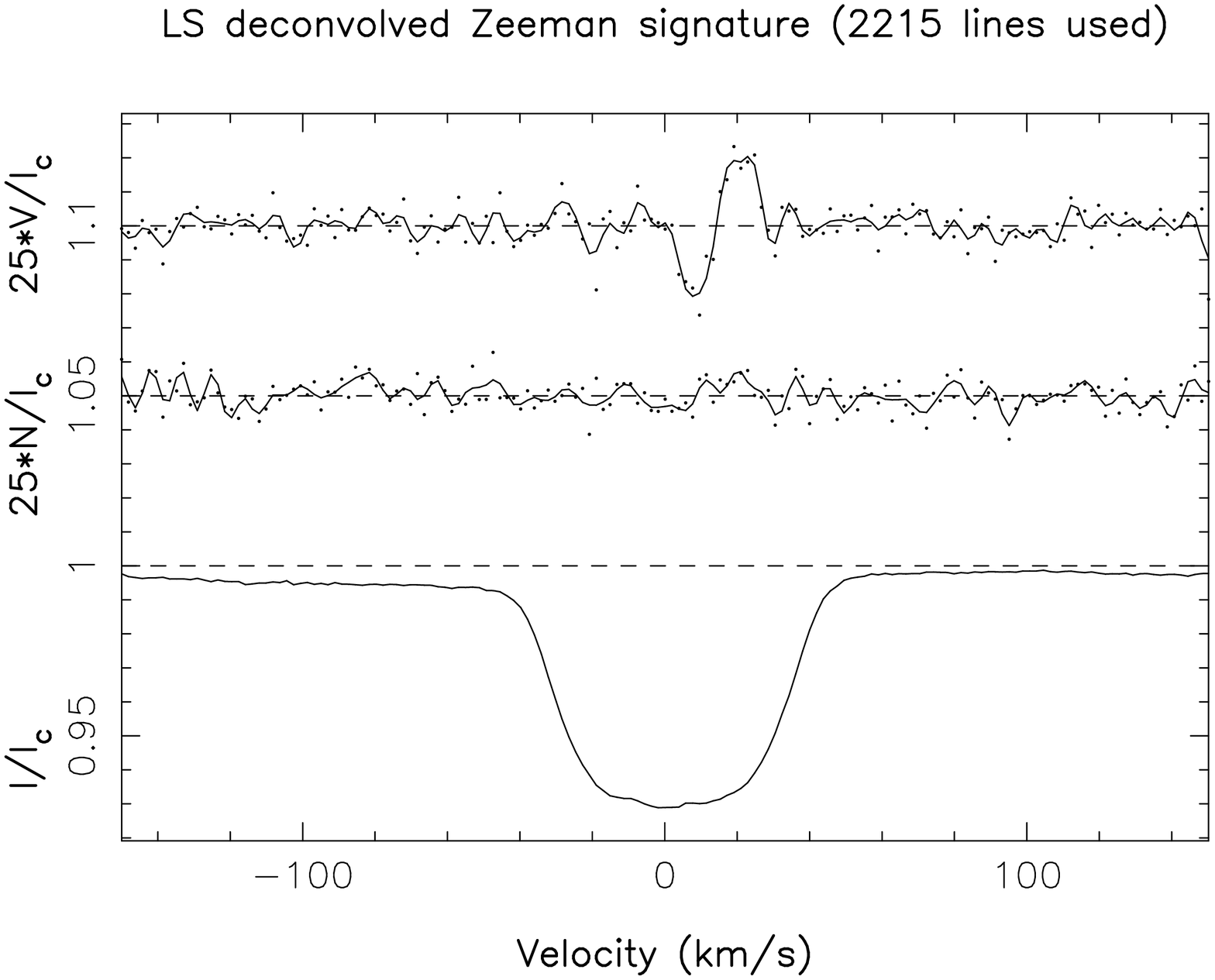}
\caption{The various LSD profiles, as explained in Figure \ref{hip21632_lsd}, for HIP~43720. The Stokes $\it{V}$ profile (upper profile) shows evidence of a strong magnetic feature on the young star HIP 43720. The LSD profile (lower profile) has a flat bottom, possibly indicating the presence of a polar spot feature.}
\label{hip43720_lsd}
\end{center}
\end{figure}

There is evidence of an active chromosphere with an emission equivalent width for the H$\alpha$ line of $\sim$400$\pm$28 m\AA\ and strong emission in the magnesium triplet lines, as shown in Figure \ref{activity_indicators}. Definite magnetic fields were observed on three occasions, however no magnetic field was detected on a fourth observation. On that occasion, the mean S/N was $\sim$ 2626. Figure \ref{hip43720_lsd} shows the Stokes $\it{V}$, null and Stokes $\it{I}$ profile. This is an interesting target for follow-up spectropolarimetric studies at the AAT and as a result, is the subject of a forthcoming intense ZDI study (Waite et al., in preparation).

\subsubsection{HIP~48770} 

\begin{figure}[ht] 
\begin{center} 
\includegraphics[scale=0.3, angle=0]{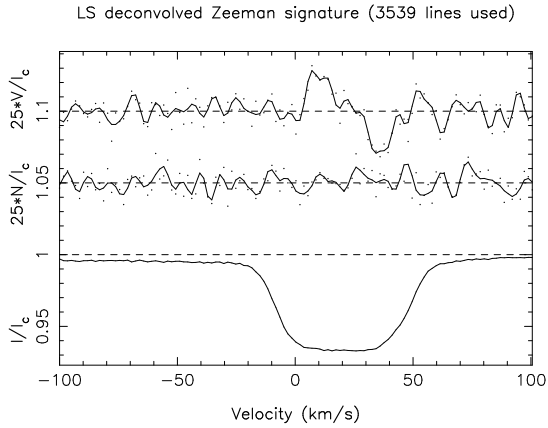}
\caption{The various LSD profiles, as explained in Figure \ref{hip21632_lsd}, for HIP 48770.}
\label{hip48770_lsd}
\end{center}
\end{figure}

HIP~48770 is a G7V pre-main-sequence star \citep{Torres06}. It has a trigonometric parallax of 5.83$\pm$1.55 mas \citep{vanLeeuwen07} giving a distance of $554_{-115}^{+199}$ ly. The \vsinis was measured to be 35 \kms, which is consistent with that found by \citet{Torres06}. The radial velocity was determined to be 19.6 \kms, which is different from the 22.6 \kmss observed by \citet{Torres06}. There appears to be no evidence of a secondary component in the spectra; but the presence of a companion in a large orbit cannot be ruled out. HIP~48770 is very young with a predominate lithium feature, as shown in Figure \ref{age_indicator}. \citet{Ammons06} calculated an effective temperature of 5539 K. However, using the bolometric corrections of \citet*{Bessell98}, our estimate is higher at 6000$\pm$300 K. It is also very active with the H$\alpha$ spectral line being almost entirely filled in; and the magnesium triplet is also very strong. A visual magnitude of 10.5 would normally make it a challenging target for SEMPOL at the AAT. However, a magnetic detection was observed at the AAT demonstrating its highly active nature. The Stokes $\it{I}$ LSD profile, along with the Stokes $\it{V}$ profile is shown in Figure \ref{hip48770_lsd}. 

\subsubsection{HIP~62517}

\begin{figure}[ht] 
\begin{center} 
\includegraphics[scale=0.3, angle=0]{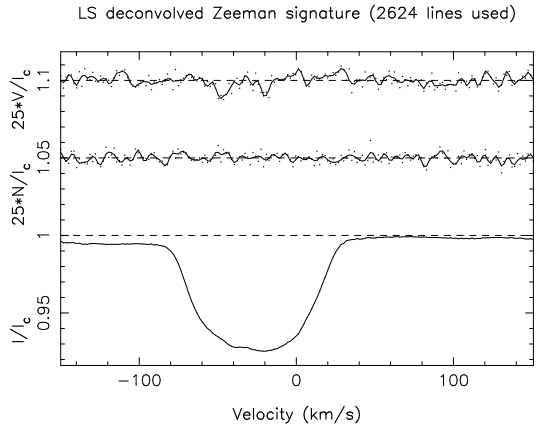}
\caption{The marginal magnetic detection in the Stokes $\it{V}$ profile (upper profile) for HIP~62517. The profiles are as explained in Figure \ref{hip21632_lsd}.}
\label{hip62517_lsd}
\end{center}
\end{figure}

HIP~62517 is an active G0 star \citep{SAO66}. It has a parallax of 2.38$\pm$1.6 mas \citep{vanLeeuwen07} giving a distance of $420_{-169}^{+862}$ ly. This star has a \vsinis of 52 \kms. This particular star has a strong H$\alpha$ emission of 265 m\AA\ coupled with some filling in of the core of the magnesium triplet. However, its lithium line is rather weak, indicating that it may not be as young as some of the other stars in the sample. \citet{Ammons06} calculated an effective temperature of 5336K, which is slightly higher than our estimate of 5250$\pm$65 K found using the bolometric corrections of \citet*{Bessell98}. A marginal detection of a magnetic field was recorded on 2010, April 2, as shown in Figure \ref{hip62517_lsd}. The S/N was 6685. Although only two snapshots were taken several months apart, with the indications that this star is single, the global magnetic field may be relatively weak and would be difficult to recover any magnetic features if observed over several epochs. This makes this star a difficult target for ZDI at the AAT. 

\subsubsection{HIP~71933} 

HIP~71933 is a pre-main-sequence F8V star \citep{Torres06}. It has a parallax of 11.91$\pm$0.99 mas \citep{vanLeeuwen07} giving a distance of $274_{-21}^{+25}$ ly. It has a projected rotational velocity of 75 \kmn. The radial velocity was measured to be $\sim$4 \kms. As mentioned previously, the large spots on the surface of this star makes accurate radial velocity measurements difficult. However, \citet{Torres06} measured 8.7 \kmss while \citet{Gontcharov06} measured 12.3$\pm$0.4 \kmss and \citet{Kharchenko09} measured 12.1$\pm$0.4 \kms. Perhaps this star is part of a wide binary system. \citet{Torres06} flagged that this star might be a spectroscopic binary star. If the star was a binary, and the companion's profile is overlapping the primary's profile, it could be mistaken for spots. Alternatively, the companion may be a faint M-dwarf star thereby not deforming the profile at all. Careful consideration of the LSD profiles produced from the high-resolution spectra obtained using the normal UCLES setup (R$\sim$50000) and spectropolarimetry (R$\sim$70000), suggest that the deformations are due to spots rather than a companion. However, the presence of a secondary component cannot be ruled out by this survey. 

\citet{Holmberg09} estimated the effective temperature to be 5900K while \citet{Ammons06} estimated the temperature to be 5938K. The equivalent width of the Li{\sc i} was measured to be 139$\pm$7 m\AA, after accounting for the Fe{\sc i} blended line using equation \ref{lithium}. When using the theoretical isochrones of \citet*{Siess00}, as shown in Figure \ref{evolution}, this star's age is estimated to be $\sim$20 Myr years and has a mass of $\sim$ 1.2 M$_\odot$.

\begin{figure}[ht] 
\begin{center}
\includegraphics[scale=0.3, angle=0]{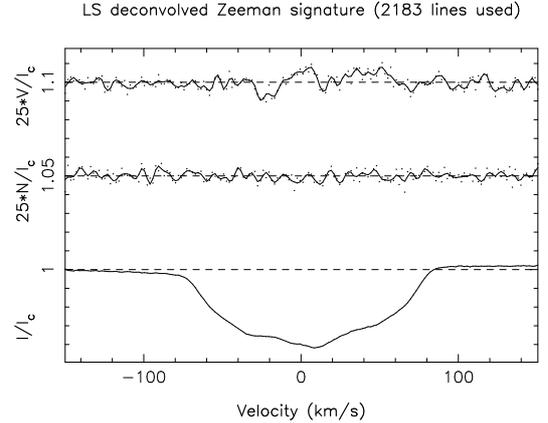}
\caption{The various LSD profiles, as explained in Figure \ref{hip21632_lsd}, for HIP 71933. The S/N for this was 8843.}
\label{hip71933_lsd}
\end{center}
\end{figure}

This is a particularly active star, as shown in Figure \ref{activity_indicators}, with emission in the magnesium triplet lines and the H$\alpha$ line. However, there was some core emission in the Na{\sc i} D$_{2}$, but not in the D$_{1}$ line. Spectropolarimetry was conducted on two occasions, once when the seeing was very poor ($\sim$ 2.5 to 3.5 arcsec) and only a S/N of 1748 but on the other occasion, reasonable seeing ($\sim$1.5 arcsec) resulted in a definite detection of a magnetic field. This detection is shown in Figure \ref{hip71933_lsd}.

\subsubsection{HIP~77144}

\begin{figure}[ht] 
\begin{center}
\includegraphics[scale=0.3, angle=0]{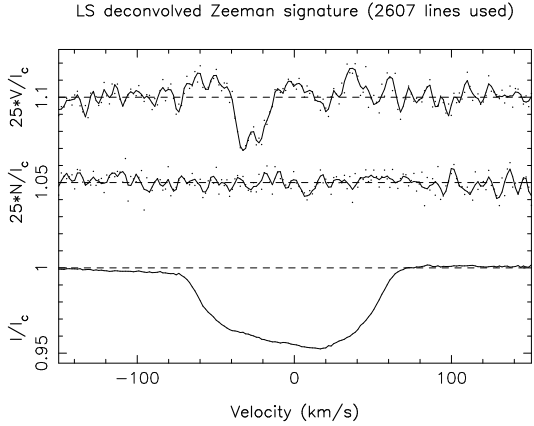}
\caption{The various LSD profiles, as explained in Figure \ref{hip21632_lsd}, for HIP~77144.}
\label{hip77144_lsd}
\end{center}
\end{figure}

HIP~77144 is a post T-Tauri G1V star \citet{Sartori03} in the Scorpius-Centaurus group \citep{Mamajek02}. It has a parallax of 7.12$\pm$1.28 mas \citep{vanLeeuwen07} giving a distance of $458_{-70}^{+100}$ ly. It has a \vsinis of 65 \kmss with a particularly strong H$\alpha$ (EEW = 466$\pm$24 m\AA) and Li {\sc I} lines (EqW = 207$\pm$7 m\AA). The temperature is estimated to be $\sim$6000$\pm$130 K. It is approximately $1.45_{-0.18}^{+0.25}$ R$_\odot$ when using the bolometric corrections of \citet*{Bessell98}, giving a luminosity of $2.4_{-0.7}^{+1.2}$ L$_\odot$. This young star, when placed on the theoretical isochrones of \citet*{Siess00} gives an age of this star of $\sim$ 20 Myr and is approximately 1.3$\pm$0.1 M$_\odot$. This is shown in Figure \ref{evolution}. The radial velocity was measured to be -1.6 \kms, which is consistent with that found by \citet{Madsen02} and \citet{Kharchenko07} to within the respective errors of each measurement. A definite magnetic field was detected on 2010, March 31, with a mean S/N in the Stokes $\it{V}$ profile of 4315. This magnetic detection is shown in Figure \ref{hip77144_lsd}.

\subsubsection{HIP~90899}     

\begin{figure}[ht] 
\begin{center}
\includegraphics[scale=0.3, angle=0]{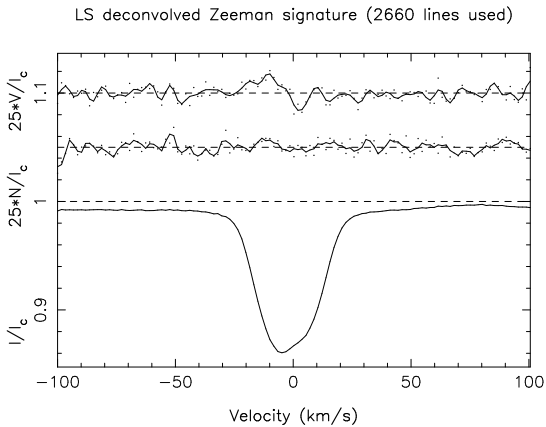}
\caption{The various LSD profiles, as explained in Figure \ref{hip21632_lsd}, for HIP~90899.}
\label{hip90899_lsd}
\end{center}
\end{figure}

HIP~90899 is a G1V star \citep{Turon93}. According to Hipparcos database, this star has a parallax of 10.87$\pm$1.34 mas \citep{vanLeeuwen07}, giving a distance of $300_{-33}^{+42}$ ly. Using the V-I determined by the Hipparcos Star Mapper Photometry, a value of 0.62$\pm$0.03 gives a temperature of $6090_{-140}^{+130}$ K using the formulation given in \citet*{Bessell98}. This is consistent with other authors such as \citet{Wright03} (6030 K) and \citet{Ammons06} (5988 K). Using the bolometric corrections of \citet*{Bessell98} gives a radius of $0.97_{-0.08}^{+0.1}$R$_\odot$ and a luminosity of $1.13_{-0.25}^{+0.37}$ L$_\odot$. This star has an emission equivalent width for the H$\alpha$ line of 408$\pm$14 m\AA, with some filling of the core of the magnesium triplet lines, meaning that it is a very active star. It has a very deep Li{\sc i} line with an equivalent width of 176$\pm$6 m\AA, suggesting that, in the absence of a companion star, is youthful in nature. The magnetic detection, as shown in Figure \ref{hip90899_lsd} was only marginal but still, at a \vsinis of 19 \kms, the magnetic topologies should be recoverable at the AAT.

\subsubsection{HIP~105388}  

\begin{figure}[ht] 
\begin{center}
\includegraphics[scale=0.3, angle=0]{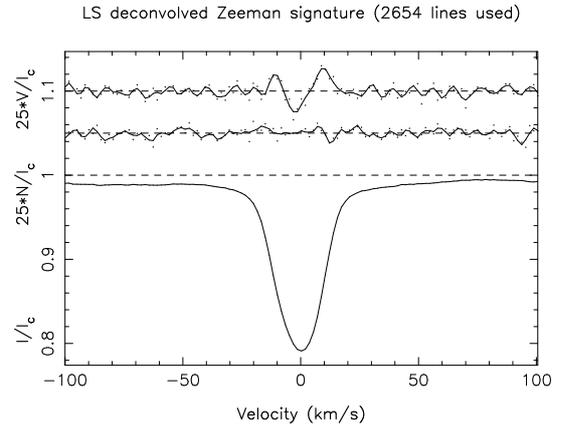}
\caption{The various LSD profiles, as explained in Figure \ref{hip21632_lsd}, for HIP~105388. The Stokes $\it{I}$ LSD profile shows limited spot activity on the star yet there is a definite magnetic detection in the Stokes $\it{V}$ profile. This is supported by the absence of signal in the null profile.}
\label{hip105388_lsd}
\end{center}
\end{figure}

HIP~105388 is a G7V pre-main-sequence star \citep{Torres06}. This star has a \vsinis of 17 \kms. This compares with the measurement of \citet{Torres06} of 15.4 \kms. The value of this star's radial velocity was determined to be -1.8$\pm$1.0 \kms. This value is consistent with those measured by \citet{Torres06} of -0.9 \kmss and \citet{Bobylev06} -1.6$\pm$0.2 \kms, to within the respective errors. This star has an emission equivalent width for the H$\alpha$ line of 520 m\AA, however, there is little core emission in either the magnesium triplet or sodium doublet. \citet{Zuckerman04} proposed that HIP~105388 was a member of the Tucana/Horologium Association and an age estimate for this moving group, hence this star, is 30 Myr. Further indication of the youthful nature of this star is the very strong Li{\sc i} line, as shown in Figure \ref{lithium}. \citet{Tetzlaff11} estimate that this is a 1.0$\pm$0.1 M$_\odot$ star. Whereas the \vsinis is at the lower limit for ZDI at the AAT, a magnetic detection on this star was secured with an S/N of only 3822. There is limited spot activity as evidenced by the smooth Stokes $\it{I}$ profile (lower LSD profile) in Figure \ref{hip105388_lsd}. Recovering magnetic features from slow to moderate rotators is possible, as shown by \citet{Petit05} and \citet{Petit08}.

\subsection{Ultra-Rapid Rotator: HIP~89829}   

\begin{figure}[ht] 
\begin{center}
\includegraphics[scale=0.3, angle=0]{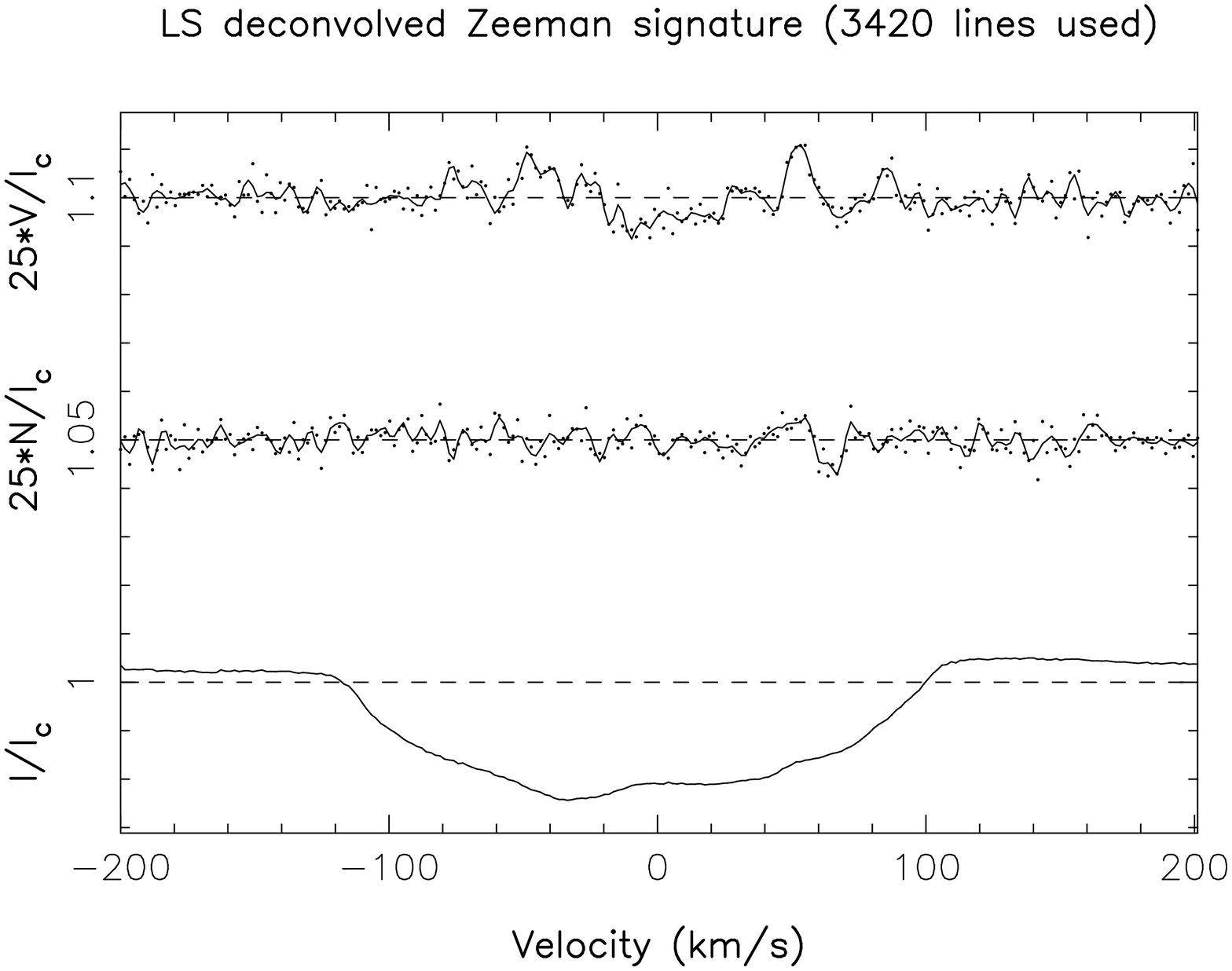}
\caption{The various LSD profiles, as explained in Figure \ref{hip21632_lsd}, for HIP~89829.}
\label{hip89829_lsd}
\end{center}
\end{figure}

HIP~89829 is a G5V star \citep{Torres06}. With a \vsinis of 114 \kms, this star has been classified as an URR. This measurement is consistent with \citet{Torres06}. \citet{Pojmanski02} quote a rotational period of 0.570751d with a photometric amplitude of $\delta$V = 0.07. This star is very active and has an emission equivalent width for the H$\alpha$ line of 280$\pm$52 m\AA\ yet little if any emission in the magnesium triplet or the sodium doublet is seen. It has a very deep, albeit broadened, lithium line with an equivalent width of 211$\pm$13 m\AA. This indicates, in the absence of a companion star, a youthful star. This is consistent with that found by \citet{Torres06}. When placed on the theoretical isochrones of \citet*{Siess00}, this star is approximately 25-30 Myr and has a mass of between 1.0 and 1.2 M$_\odot$. The magnetic detection is shown in Figure \ref{hip89829_lsd}. This is one of the most rapidly rotating stars that has had its magnetic field detected at the AAT. 

As mentioned in Section \ref{vrad}, the large variation seen in the radial velocity measurements for HIP~89829 could be a result of this star being a binary star. However, after carefully considering the resulting LSD profiles from both the normal UCLES setup (R$\sim$50000) and SEMPOL setup (R$\sim$70000), we feel that this star is {\bf probably} single. 

\subsection{Hyper-Rapid Rotator: HIP~93378}

HIP~93378 is a pre-main-sequence, G5V star \citep{Torres06}. According to Hipparcos database, this star has a parallax of 9.14$\pm$0.92 mas \citep{vanLeeuwen07}, giving a distance of $357_{-33}^{+40}$ ly. It is a HRR with a \vsinis of 226 \kms, which is consistent with the 230 \kmss value measured by \citet{Torres06} within the large uncertainty created by such a rapid rotation. This star's H$\alpha$ profile, when matched against a rotationally broadened solar profile, exhibitesd no emission in the core. Thus it appears that there is limited chromospheric activity occurring on this star. However, as mentioned previously, this lack of chromospheric emission may be due to the extreme broadening of the spectral lines thereby ``washing out" the emission component. Another reason for this decreased H$\alpha$ emission may be due to supersatuation or even a modification of the chromospheric structure by the extremely strongs shear forces as a result of such rapid rotation.

\begin{figure}[ht] 
\begin{center}
\includegraphics[scale=0.3, angle=0]{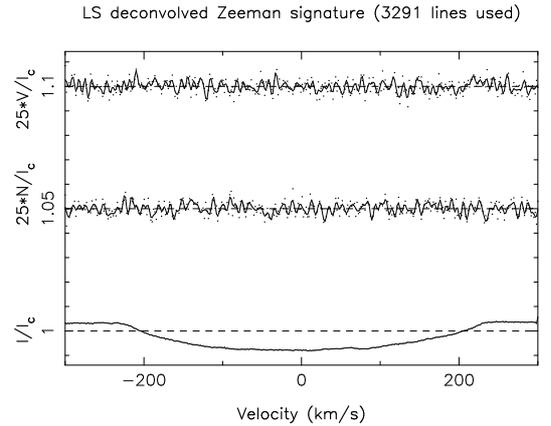}
\caption{The various LSD profiles, as explained in Figure \ref{hip21632_lsd}, for HIP~93378. }
\label{hip93378_lsd}
\end{center}
\end{figure}

No magnetic detection was observed on this star, even when the data were binned to increase the relative signal-to-noise in excess of 9000. Again this extreme rotation may have simply washed out any magnetic signature. The resulting LSD profile is shown in Figure \ref{hip93378_lsd}. Although this star is an extremely difficult target for ZDI at the AAT, its extreme rotation makes it an interesting target for Doppler imaging.

\subsection{Active, young, slowly rotating stars.}

\begin{figure}[ht] 
\begin{center}
\includegraphics[scale=0.35, angle=0]{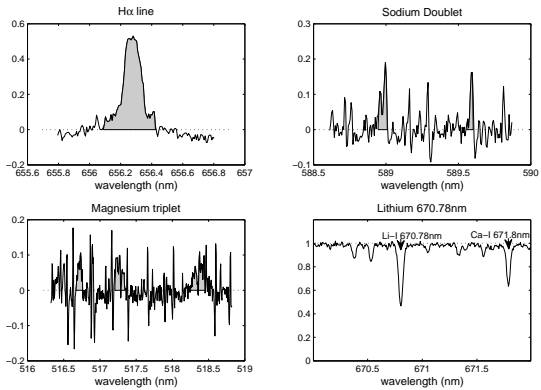}
\caption{The activity indicators for HIP~68328. This star has a very strong emission component of the H$\alpha$ line with some moderate filling in of the sodium doublet and magnesium triplet. Also shown in this figure is the strong Li{\sc i} 670.78 nm line.}
\label{hip68328_activity}
\end{center}
\end{figure}

This survey also found a number of slower rotating stars that are very young. Due to time constraints we did not take spectropolarimetric observations of these stars. However, the CFHT and TBL have been able to recover magnetic fields on slow rotators \citep{Petit08}. One such star is the early G0 \citep{Sartori03} star HIP~68328. It has a relatively slow rotation of $\sim$ 6 \kms. It has a parallax of 8.34$\pm$1.56 mas \citep{vanLeeuwen07} giving a distance of $391_{-61}^{+90}$ ly. Using the bolometric corrections of \citet*{Bessell98}, the temperature of HIP 68328 was estimated to be $5420_{-107}^{+114}$ K. This is consistent with the temperature estimated by \citet{Lafrasse10} even though it is lower than the most recent estimate by \citet{BailerJones11} of 5871 K. This project estimates that this star has a radius of $1.4_{-0.17}^{+0.26}$ R$_\odot$, giving a luminosity of $1.49_{-0.44}^{+0.78}$ L$_\odot$. This value is similar, within the relative error bars, to that estimated by \citet{Sartori03} of 1.66 L$_\odot$. \citet{DeZeeuw99} identified this star as a possible member of the Scorpius-Centaurus OB association. This is a young star-forming region with stars less than 20 Myr old. When placing this star on the theoretical isochrones of \citet*{Siess00}, as shown in Figure \ref{evolution}, the star has an age of $\sim$20$\pm$10 Myr and a mass of $\sim$1.2$\pm$0.1 M$_\odot$. This age is consistent with the observation of a very deep Li{\sc i} line with an equivalent width of 263$\pm$4 m\AA. It has a emission equivalent width for the H$\alpha$ of 850$\pm$45 m\AA, the most active star in this sample by this measure. These observations of youth and activity are shown in Figure \ref{hip68328_activity}. Other slow rotators with substantial lithium lines include HIP~23316, HIP~41688, HIP~63734 and HIP~11241. All have indicators of having an active chromosphere with core emission in the H$\alpha$ line. Another interesting target is the slow rotator is HIP~5617. It has very prominent emission in the wings of the H$\alpha$ line, extending above the continuum, indicating the presence of circumstellar material. However, the lithium line is very weak. This is perhaps a T-Tauri star.

\subsection{Binary and Multiple Stellar Systems}
\begin{figure}[ht] 
\begin{center}
\includegraphics[scale=0.45, angle=0]{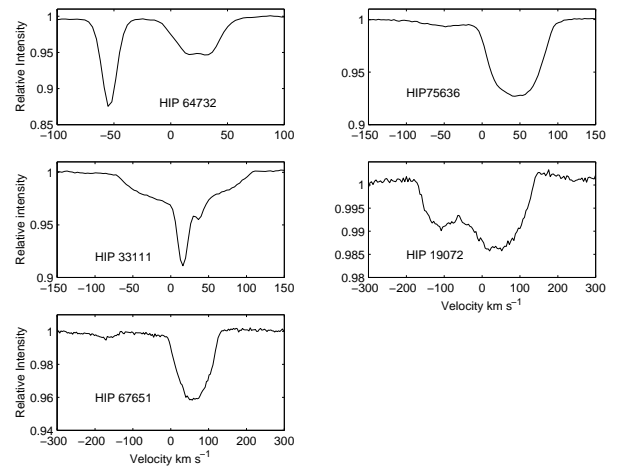}
\caption{LSD profiles of some likely binary stars.}
\label{binary}
\end{center}
\end{figure}

The LSD profile is an excellent way of identifying a binary star \citep{Waite05}, as the LSD profiles often show both stars, except if the star is undergoing an eclipse or is a faint star such as an M-dwarf. Where stars that appear to be rapidly rotating, more than one spectra (often two, three or more) were taken to make sure that the star was indeed single. HIP~31021, HIP~64732 and HIP~73780 were identified as binary stars based on their individual LSD profiles. HIP~64732 may have a giant polar spot on one of the components as one of the LSD profiles exhibited a ``flat bottom", indicating the likely presence of a giant polar spot or high latitude features. This is shown in top left panel of Figure \ref{binary}. HIP~19072, HIP~67651 and HIP~75636 are likely to be spectroscopic binary stars while HIP~33111 could be a triple system. The associated LSD profiles are shown in Figure \ref{binary}. There is a slight deformation of the LSD profile of HIP~75636 on the blue side that could be due to a companion star just moving into an eclipse of the second star. Also, the radial velocity of this star was measured to be 43.7$\pm$1 \kmss whereas \citet{Torres06} measured the radial velocity to be 5.9 \kms. While some of the stars in this survey exhibit slight shifts in the radial velocity when compared with the work of \citet{Torres06}, this difference is far too great to conclude anything else except that it is a binary star.  

Table \ref{summary} gives a summary of the likely targets for follow-up magnetic studies.

\begin{table*}
\begin{center}
\caption{New solar-type targets for Zeeman Doppler imaging.}
\label{summary}
\begin{tabular}{ll}
\hline	
Confirmed targets  & HIP number              \\
\hline
MR 		      & 21632, 90899, 105388     \\
RR 		      & 43720, 48770, 62517, 71933, 77144    \\
URR		      & 89829                   \\
\hline
Potential targets ${^1}$ & \\
\hline
SR \& MR	     & 5617, 10699, 11241, 23316, 25848,  \\
		     & 63734, 68328    \\
HRR		     & 93378 (DI target)              \\
\hline
Binary Stars & 					\\
\hline
Binary Stars &	31021, 64732, 73780   \\
Probable Binary Stars   & 19072, 33111${^2}$, 67651, 75636   \\
\hline
\end{tabular}
\medskip\\
See Table \ref{vsini_def} for definitions used in this table.\\
${^1}$ Based solely on activity (if SR or MR) or rapid rotation. \\
${^2}$ This star may even well be a triple system. \\ 
\end{center}
\end{table*}

\section{Conclusion}

This survey aimed to determine the nature of a number of some unresolved variable stars from the Hipparcos database and to identify a number of targets for follow-up spectropolarimetric studies at the AAT. Of the 38 stars observed, three stars (HIP~31021, HIP~64732, HIP~73780) were spectroscopic binary stars while further three stars, (HIP~19072, HIP~67651 and HIP~75636) are likely to be a spectroscopic binary stars while HIP~33111 could be a triple system. Two stars rotate with speeds in excess of 100 \kms: HIP~93378 (\vsinis $\sim$ 226 \kms) and HIP~89829 (\vsinis $\sim$ 114 \kms). Magnetic fields were detected on a number of the survey stars: HIP~21632, HIP~43720, HIP~48770, HIP~62517, HIP~71933, HIP~77144, HIP~89829, HIP~90899 and HIP~105388. All of these stars would be suitable for follow-up spectropolarimetric studies using SEMPOL at the AAT.



\section*{Acknowledgments} 

The authors thanks the Director of the Australian Astronomical Observatory for allowing us some of his observing time to observe potential targets. This time will permit the expansion of our study into activity cycles on solar-type stars. We thank the referee of this paper, Pascal Petit, for his diligence and insightful comments that has made this a better paper. We thank J.-F Donati on supplying ES{\sc p}RIT. This project is supported by the Commonwealth of Australia under the International Science Linkages program. This project used the facilities of SIMBAD, HIPPARCOS and IRAF. This research has made use of NASA's Astrophysics Data System.


\end{document}